# Energy spectra of abundant cosmic-ray nuclei in the NUCLEON experiment

N.Gorbunov[ab], V.Grebenyuk[ab], D.Karmanov[c], I.Kovalev[c], I.Kudryashov[c], A.Kurganov[c], A.Panov[c], D.Podorozhny[c], S.Porokhovoy[a], L.Sveshnikova[c], A.Tkachenko[ad], L.Tkachev[ab], A.Turundaevskiy[c], O.Vasiliev[c], A.Voronin[c]

[a] *Joint Institute for Nuclear Research, Dubna, 141980, Russia*
[b] *Dubna State University, Dubna, 141980, Russia*
[c] *Skobeltsyn Institute of Nuclear Physics, Moscow State University, Moscow, 119991, Russia*
[d] *Bogolubov Institute for Theoretical Physics, Kiev, 03680, Ukraine*

**Abstract**

The NUCLEON satellite experiment is designed to directly investigate the energy spectra of cosmic-ray nuclei and the chemical composition (Z=1−30) in the energy range of 2–500 TeV. The experimental results are presented, including the energy spectra of different abundant nuclei measured using the new Kinematic Lightweight Energy Meter (KLEM) technique. The primary energy is reconstructed by registration of spatial density of the secondary particles. The particles are generated by the first hadronic inelastic interaction in a carbon target. Then additional particles are produced in a thin tungsten converter, by electromagnetic and hadronic interactions.

**1. Introduction**

The "knee" energy range - $10^{14}$ - $10^{16}$ eV - is a crucial region for the understanding of cosmic rays, acceleration and propagation in the interstellar medium. It is important to obtain more data with elemental resolution.

There are no direct measurements of cosmic ray nuclei spectra in the "knee" energy range. The main information about cosmic ray nuclei at $10^{12}$ - $10^{14}$ eV has been obtained by balloon (ATIC[1,2], CREAM [3,4], TRACER [5]) and satellite (AMS02 [6,7] for lower energies, SOKOL [8]) experiments. The CALET experiment [9] is performed onboard ISS now. The DAMPE experiment [10] has also been realised. However, additional direct measurements at energies of up to 1000 TeV are necessary.

The NUCLEON satellite experiment is designed to directly investigate, above the atmosphere, the energy spectra of cosmic-ray nuclei and the chemical composition from 2 to more than 500 TeV (before the "knee"). The highest measured energy is equal to 900 TeV.

2. **The NUCLEON design**

The NUCLEON device [11-16] was designed and produced by the collaboration of SINP MSU (the main investigator), JINR (Dubna) and a number of other Russian scientific and industrial centres. Currently, it is placed on board the RESURS-P №2 satellite. The spacecraft's orbit is a Sun-synchronous one, with an inclination of 97.276° and a middle altitude of 475 km. The satellite was launched on 26 December, 2014.

Scientific objectives and detection techniques determined the detector design. The general composition of the NUCLEON apparatus is presented in fig. 1.

The new Kinematic Lightweight Energy Meter (KLEM) technique was applied. The primary energy is reconstructed by registration of spatial density of the secondary particles. The particles are generated by the first hadronic inelastic interaction in a carbon target. The equivalent thickness of the carbon target is equal to 0.23 proton interaction lengths.

Additional particles are produced in thin tungsten converter by electromagnetic and hadronic interactions. The spatial density of the secondary particles is measured by silicon microstrip detectors.

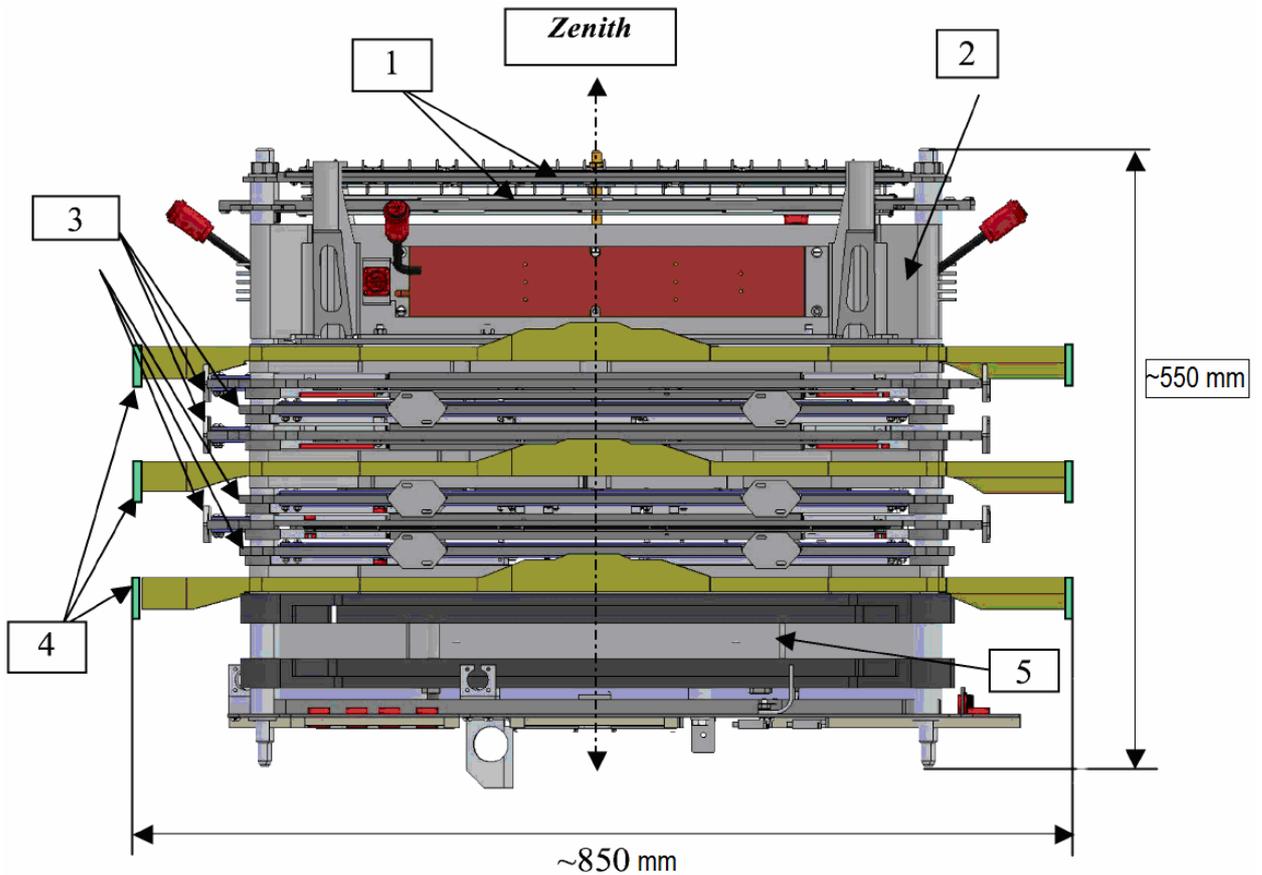

Figure 1: Simplifed layout of NUCLEON experiment scientifc equipment. (1) - two pairs of charge measurement system planes; (2) - carbon target; (3) - 6 planes of energy measurement system utilizing the KLEM technique; (4) - 3 double trigger system planes; (5) - calorimeter.

The NUCLEON apparatus includes different units based on silicon and scintillator detectors. The charge measurement system consists of 4 pad silicon detectors layers. The KLEM energy measurement system includes 6 silicon microstrip detectors interleaved with thin tungsten layers. There are six layers of plastic scintillator detectors of 0.75 cm thickness in the trigger system. They generate necessary trigger signals for the KLEM system. Each of the trigger planes consists of 16 scintillator strips. The light signals from the strips are collected by WLS fibres to PMTs. The calorimeter also includes 6 silicon microstrip detectors.

The total weight of the device is about 375 kg. Power consumption is less than 175 W. The dimensions of the instrument are as follows: length, 85 cm; width, 85 cm ; height, 55 cm.

The effective geometric factor is more than 0.2 m²sr for the KLEM system and nearly 0.06 m²sr for the calorimeter. The surface area of the device is equal to 0.25 m². The charge measurement system provides a resolution of 0.15–0.20 charge units.

The set of data obtained by all detectors can be considered as the image of the event. An event example is presented in fig. 2. The reconstructed trajectory crosses charge detectors (1), KLEM system silicon microstrip detectors (3), and calorimeter silicon microstrip detectors (5). We can see a projection of the cascade in the device.

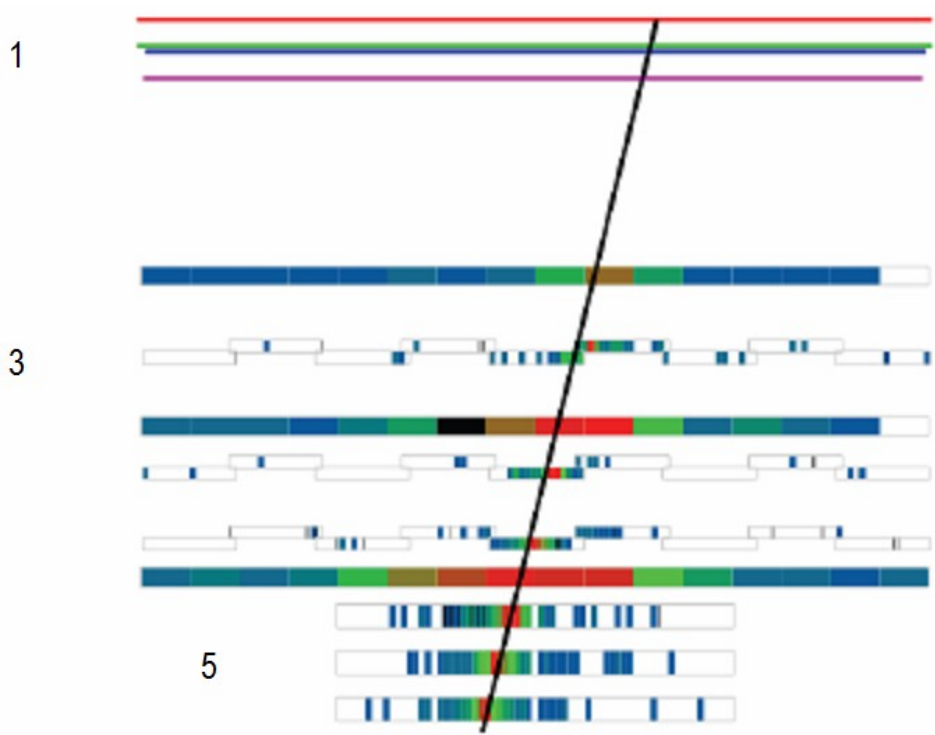

Figure 2. The image of the event. The nucleus initialized the shower. (1) - two pairs of charge measurement system planes; (3) - 6 planes of energy measurement system utilizing the KLEM technique; (5) - calorimeter.

## 3. The KLEM technique

A new energy measurement method KLEM was proposed in [17-21]. The technique can be used over a wide range of energies ($10^{11}$–$10^{16}$ eV) and gives an energy resolution of 70% or better, according to simulation results [13,15,16,22].

We have considerably improved a kinematical method. The simple kinematical method for the determination of the primary particle energy proposed by Castagnoli [23] gives large errors between 100% and 200%. To overcome this problem, a combined method was proposed. This is based, on the one hand, on the measurements of spatial density of not only charged secondaries, but also neutral ones. On the other hand, a measuring technique at the data processing stage is proposed, which allows for an increased contribution of faster secondaries to the energy determination, and eliminates that of slower ones.

The primary energy is reconstructed by registration of spatial density of the secondary particles. The new particles are generated by the first hadronic inelastic interaction in a carbon target. Then, additional particles are produced in the thin tungsten converter by electromagnetic and hadronic interactions. The main difference between the proposed KLEM method and ionisation calorimeters is that the KLEM technique does not need heavy absorbers for the shower measurement. Thus, it is possible to design relatively light cosmic-ray detectors with a large geometric factor.

We use the S-estimator for the energy determination:

$$S = \sum_{i=1}^{N} \eta_i^2$$

where summation is over all *N*-secondaries detected after the converter, $\eta_i = -\ln \mathrm{tg}\theta_i/2 \approx -\ln(r_i/2H)$, $\theta_i$ is the angle between the shower axis and secondary particle direction, $r_i$ is the distance from the shower axis for all position-sensitive detector layers located after the converter, and *H* is the distance from the interaction point in the target. The estimator depends on the angular distribution of secondaries. The number of secondaries with minimal angles is more sensitive to the Lorentz factor of the primary particle than the total multiplicity. The squaring increases the contribution of these particles. For the real apparatus we apply $H$ determined as the distance from the middle of the target to the detector layer. The systematic uncertainty on the energy scale inferred by this approximation is small in comparison to the shower fluctuations.

For example, the value of $\ln^2(2H/x)$ is equal to 48.4 for the middle of the target ($H$=255 mm) and 45.4 for the low boundary of the target ($H$=204 mm). Thus, the maximal systematic deviation is near 6%. The direct simulation shows that the rms deviation of reconstructed energy (see below) increased from 70% (for true interaction point) to 70.2%. On the one hand, the S-estimator characterises the distribution of secondaries on emission angles as being sensitive to the Lorentz-factor of the primary particle. On the other hand, S is proportional to the multiplicity of secondaries produced in the target and multiplied in the converter. The contribution of slow neutrals is eliminated by the squaring of $\eta$. Simulations showed the simple semi-empirical power law energy dependence for S [20].

The perpendicular projections $x_i$ and $y_i$ can be used instead of the distance $r_i$. This allows us to exploit microstrip silicon detectors for spatial measurements. The microstrip detectors can register many charged particles per strip. The signal is proportional to the strip ionisation or the number of single-charged particles. Thus, the S-estimator is defined as:

$$S = \Sigma I_k \ln^2(2H/x_k)$$

where $x_k$ is a distance between the shower axis and the strip $k$, $I_k$ is a signal in the strip $k$.

The simple semi-empirical power law $<S(E)> \sim E^{0.7-0.8}$ dependence of the energy per nucleon was obtained.

The above-mentioned squaring and multiplication of secondaries in the converter make energy dependence steeper than for multiplicity in the first interaction. For incident nucleus with mass number A not all of the nucleons interact with the target carbon nucleus. Therefore, the multiplicity of secondaries is not proportional to $A$ but the angular distribution of secondaries is similar to the distribution for one proton. The $<S(E)>$ dependence is similar for different types of primary nuclei in the wide range.

## 4. Simulations

### 4.1. The simulation of the energy measurement system.

Isotropic fluxes of protons, and helium, carbon, sulphur and iron nuclei were simulated. For constant statistical accuracy on all studied energy ranges (100 GeV — 1000 TeV), a uniform logarithmic distribution on energy was simulated dN/d(lnE)=const. The signal in the scintillator and silicon detectors was considered proportional to the energy deposit in the corresponding volume. Algorithms are completely identical to processing of the simulated and experimental databanks. Selection by trigger conditions and reconstruction of the primary particle track were reproduced. For selected events, an optimisation of the KLEM technique was performed. The calibration dependencies were calculated.

The practical applicability of the proposed KLEM energy measurement technique was estimated using the results of the simulation, employing the GEANT 3.21 [24] software package complemented by the QGSJET [25, 26] nuclear interaction generator to describe high-energy hadron–nucleus and nucleus–nucleus interactions.

### 4.2. Calibration curves for different components

For the results of the simulation analysis the following main assumptions were applied.

First, a power-law dependence of the reconstructed energy on the estimator S, defined earlier, was assumed.

$$E_{rec} = aS^b$$

Second, the distribution function on the reconstructed energy does not depend on primary energy, only on the ratio of the reconstructed and primary energy $F(E_{rec}/E)$.

Let us designate $k = E_{rec}/E$

We know that cosmic-ray spectra are close to power law.

$$\frac{dN}{dE} = AE^{-(\gamma+1)}$$

At the given reconstructed energy:

$$E = E_{rec}/k$$

$<k> = 1$ for the power energy spectrum.

Therefore, the following equations are obtained:

$$\frac{dN}{dk} = AE_{rec}^{-\gamma}k^{\gamma}$$

$$<k> = \frac{\sum k_i^{\gamma+1}}{\sum k_i^{\gamma}} = 1$$

We obtained for a simulated event with energy $E_i$

$$k_i = \frac{aS_i^b}{E_i}$$

As a result, we receive the formula for $a$:

$$a = \frac{\sum (S_i^b/E_i)^{\gamma}}{\sum (S_i^b/E_i)^{\gamma+1}}$$

By means of the ordinary least squares method, the $b$ values for different components were received from simulated databanks. The $a$ values were calculated according to the formula received above. It is necessary to calculate the unbiased value of energy for the power spectrum. In practice, the more convenient parameter, $a_2$, was applied:

$$E_{rec} = a_2(S \cdot 10^{-5})^b$$

The values of $a_2$ and $b$ are presented in Table 1.

Table 1. Calibration parameters

| Projectile nucleus | $a_2$ GeV | $b$ |
|---|---|---|
| p | 1651 | 1.36 |
| He | 2556 | 1.27 |
| C | 3514 | 1.18 |
| S | 4163 | 1.14 |
| Fe | 4362 | 1.12 |

The simulation results (S vs E) are presented in fig.3 for protons and carbon nuclei.

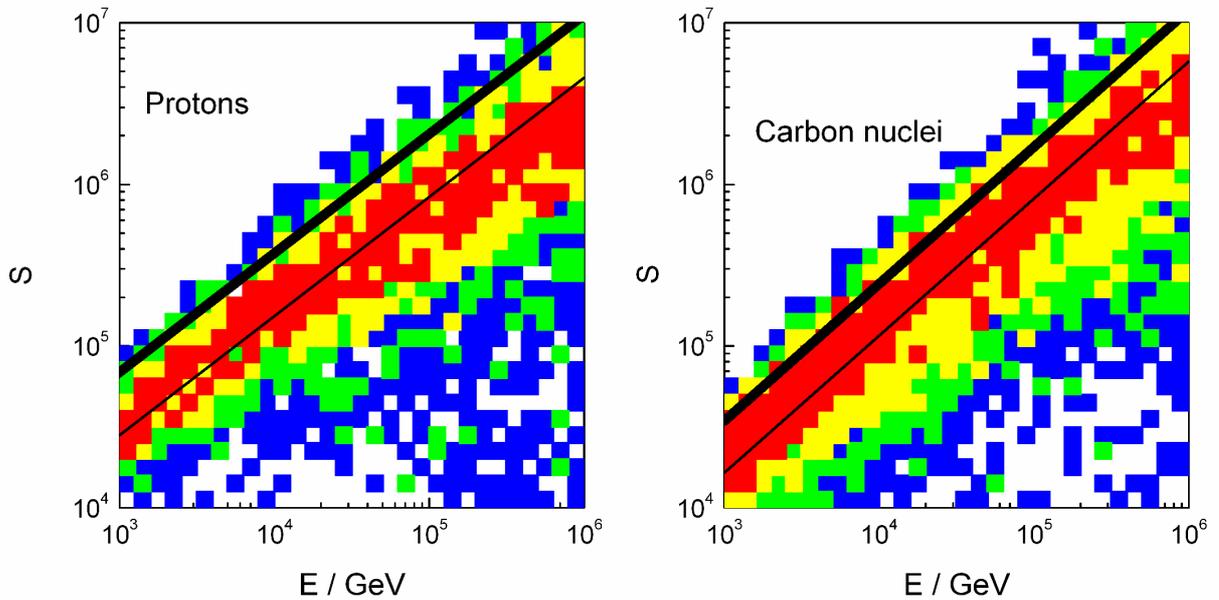

Figure 3. Simulation results. S vs E for protons and carbon nuclei. The fit (thick line) is shifted according to the power spectrum ($\gamma = 1.6$). The thin line corresponds to simple fit of data.

The primary particles were generated with uniform energy distribution in logarithmic scale by the simulation. The thin line in fig.3 is the power fit of these data. However the real cosmic-ray energy spectra are close to power law spectra.

Thus it is necessary to apply event weights according to expected energy spectra. The thick line corresponds to the fit obtained with event weights for the power spectrum.

The dependence of the energy resolution on energy is shown in fig.4 for different projectile nuclei according to parameters from Table 1, taking into account power shape of cosmic-ray spectra. The reconstructed energy distribution is a convolution of distributions at fixed energies and shape of spectrum. Thus, the resolution can depend on a spectral exponent.

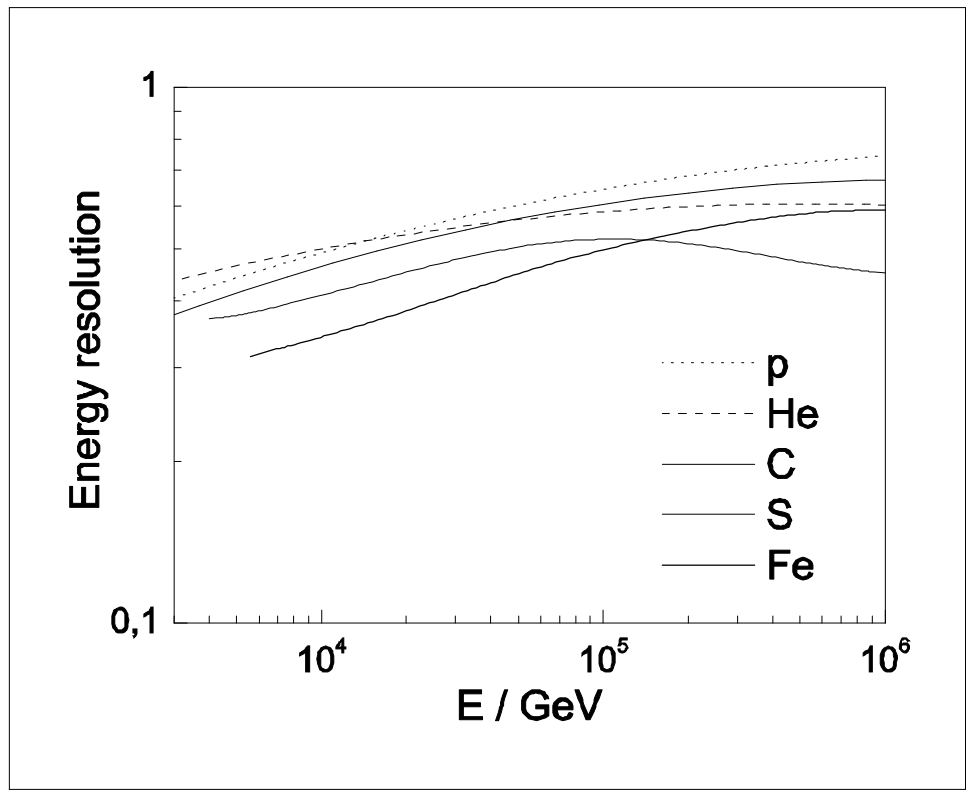

Figure 4. Energy resolution dependence on energy (simulation)

## 4.3. The registration efficiency for different nuclei

At the given algorithm of energy reconstruction, the registration efficiency can be determined as the ratio of reconstructed and primary energy spectra by simulation results. In practice, it is a combined parameter. Different effects influencing the registration efficiency were not determined separately because they were automatically taken into account by the simulation.

The registration efficiency depends on the probability of interaction of the corresponding particle in the device, trigger conditions, and also error in the energy measurements. Accounting for energy dependence of registration efficiency is necessary for the solution of the problem of deconvolution, i.e. a correct reconstruction of energy spectra.

At low energies, fragments of primary, heavy nuclei interacting in the carbon target can imitate a particle with energy near 1-3 TeV. Moreover, there are other threshold effects. It allows for reconstruction of energy spectra only at energies more than 3-5 TeV. The energy dependences of the registration efficiency calculated by simulation results, according to the definition given above, are presented in fig. 5. The registration efficiencies for light nuclei are low because of the small interaction probability in the carbon target and low multiplicity of secondary particles generated by the first inelastic interaction.The calculated dependences are rather smooth. It allows the use of interpolation of energy and a charge for various nuclei of cosmic rays.

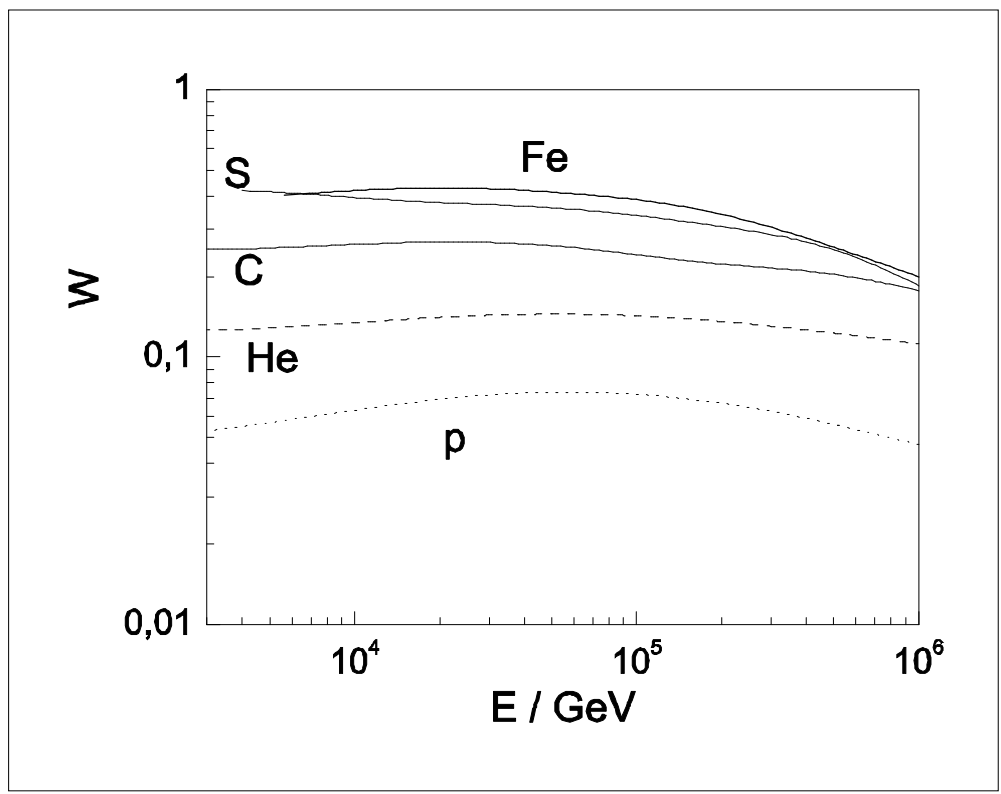

Figure 5. The registration efficency

## 5. Beam tests of the energy measurement system.

Tests of the KLEM method with NUCLEON prototypes were performed at the pion 100–350 GeV beams of the SPS accelerator at CERN [13,15]. The procedure for reconstructing the particle energy uses the theoretical calibration dependence of parameter $S(E)$. The calibration cureves for high energy protons and carbon nuclei are presented in Fig.3. Based on the simulation data, the calibration dependence $<S(E)>$ estimator of energy was plotted and the power law index was found to be 0.75.

In a second step, the NUCLEON flight model was tested. Pion data were obtained for 150 and 350 GeV. The coordinates of the shower axes were determined by the microstrip detector signals. The S values were calculated for every selected event.

Energy dependence S(E) is presented in fig.6. The data were obtained for the pion beam of 2013 (squares) and the previous tests in 2008 (circles) [11]. The curve was obtained by simulation. The point at 200 GeV is significantly off fitted line due to different trigger selection.

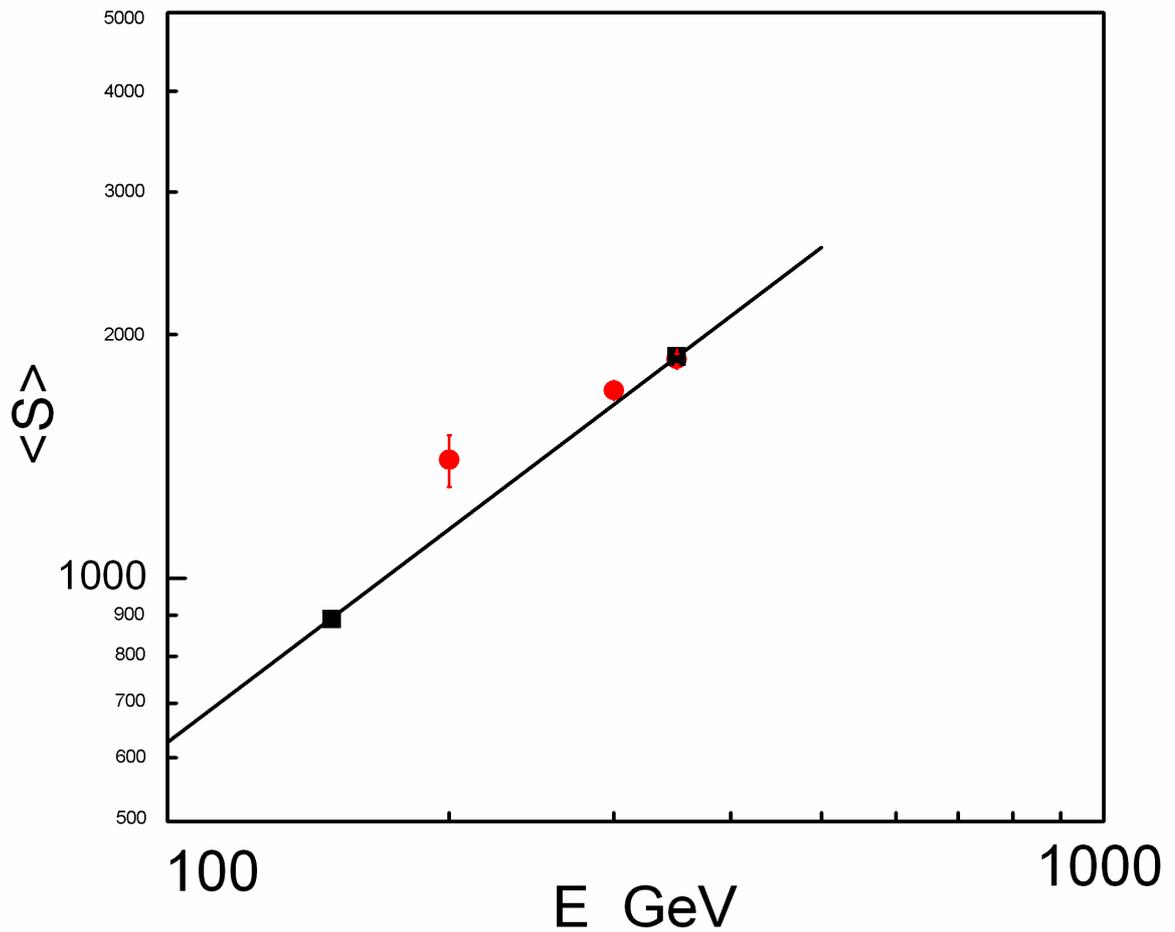

Figure 6. The calibration energy dependence S(E) for pions

The normalised distributions of the reconstructed energy for primary pions with energies of 150 (thin line) and 350 GeV (thick line) are shown in fig. 7 [11]. The rms deviation to primary energy ratio is equal to 0.53 for 150 GeV and 0.63 for 350 GeV. The asymmetry of the distributions is determined by the asymmetry of multiplicity distributions for hadron interactions.

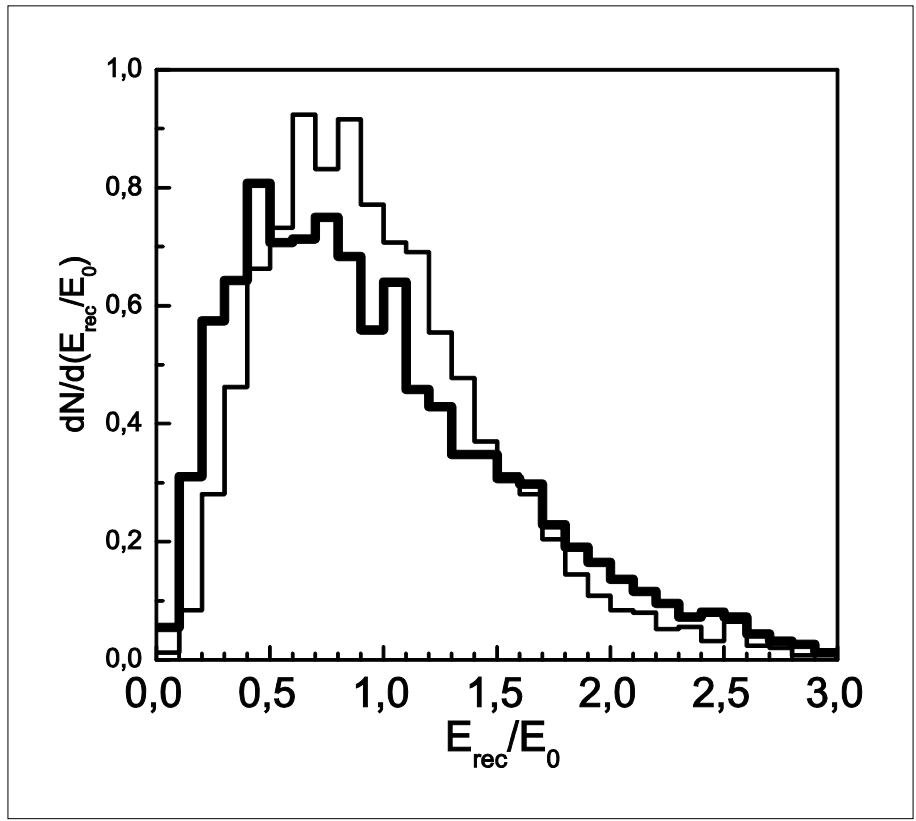

Figure 7. Normalized reconstructed energy distributions for pion beams of 150 GeV (thin line) and 350 GeV (thick line) [11].

## 6. The space experiment results
### 6.1. Charge measurements

Processing of data obtained by the satellite experiment consists of several stages.

At the first stage, the trajectory of the particle is analysed. Maxima of the spatial distribution of ionisation are searched for each microstrip detector layer of the energy measurement system. It is supposed that these maxima correspond to track cross points with the detector layers. On these points the axis is reconstructed by Ordinary Least Squares. Events where the axis is located within a fiducial acceptance are selected.

For each layer of the charge measurement system the cross point with this particle trajectory is reconstructed. Around this point, (taking into account possible error), coordinates of pads through which a primary particle can pass are determined. The signals from these pads are compared and the largest amplitude is selected.

Calibrations are necessary for transition from the registered signal amplitude to the charge value. At the first stage, the beam tests results were used. Changes for each channel were considered by means of on-board calibration.

The initial calibrations were applied to obtain the first level charge distributions with charge errors near 0.3-0.5. Further analysis showed that the dispersion of characteristics of the various detectors of the charge measurement system is the main reason for a high error.

After the registration of rather large statistics (half a year of the experiment), charge distributions were successfully received separately for each of 256 charge measurement system detectors.

For each such distribution, the reference peaks from the most abundant nuclei, e.g. protons, helium, carbon, oxygen and iron are allocated (Z=1,2,6,8,26). The additional calibration for each detector is based on these peak values. Thus, the charge distributions with high resolution (0.15-0.20 for different nuclei) were obtained (fig.8). The small offset of the charge peaks (Z>14) is caused by nonlinearity of electronics. This effect was taken into account by elemental selection.

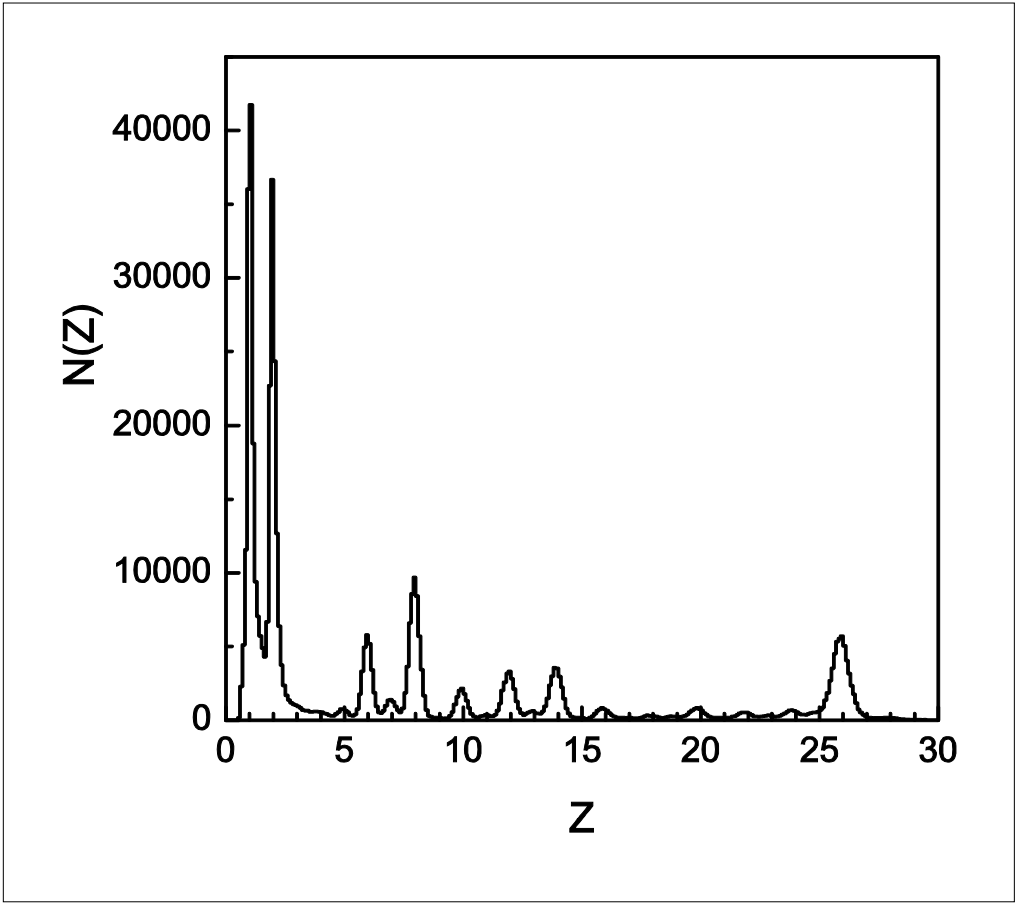

Figure 8. Charge distribution

## 6.2. Energy spectra reconstruction

By deconvolution, both energy dependence of the S-estimator and energy dependence of the registration efficiency are considered.

The differential flux of particle species is given by

$$\frac{dN}{d\ln E} = \frac{1}{\Gamma w W} \frac{\Delta N}{\Delta(\ln E)\Delta T}$$

In this formula, $\Gamma$ is the geometrical factor, w is fraction of the live time, W is the registration efficiency, $\Delta N$ is the number of registered nuclei, $\Delta(\ln E)$ is the bin width, $\Delta T$ is the exposure time.

The $\Gamma$ and W parameters were determined by Monte Carlo simulations (see 4.3). Nonlinear function E(S) allows to use for deconvolution simple diagonal matrix with $\Gamma W$ matrix element for every energy bins.

Examples of artificial spectra simulated and reconstructed by the above described method energy spectra are presented in fig.9. Both simple power spectrum and a spectrum with a break can be reconstructed.

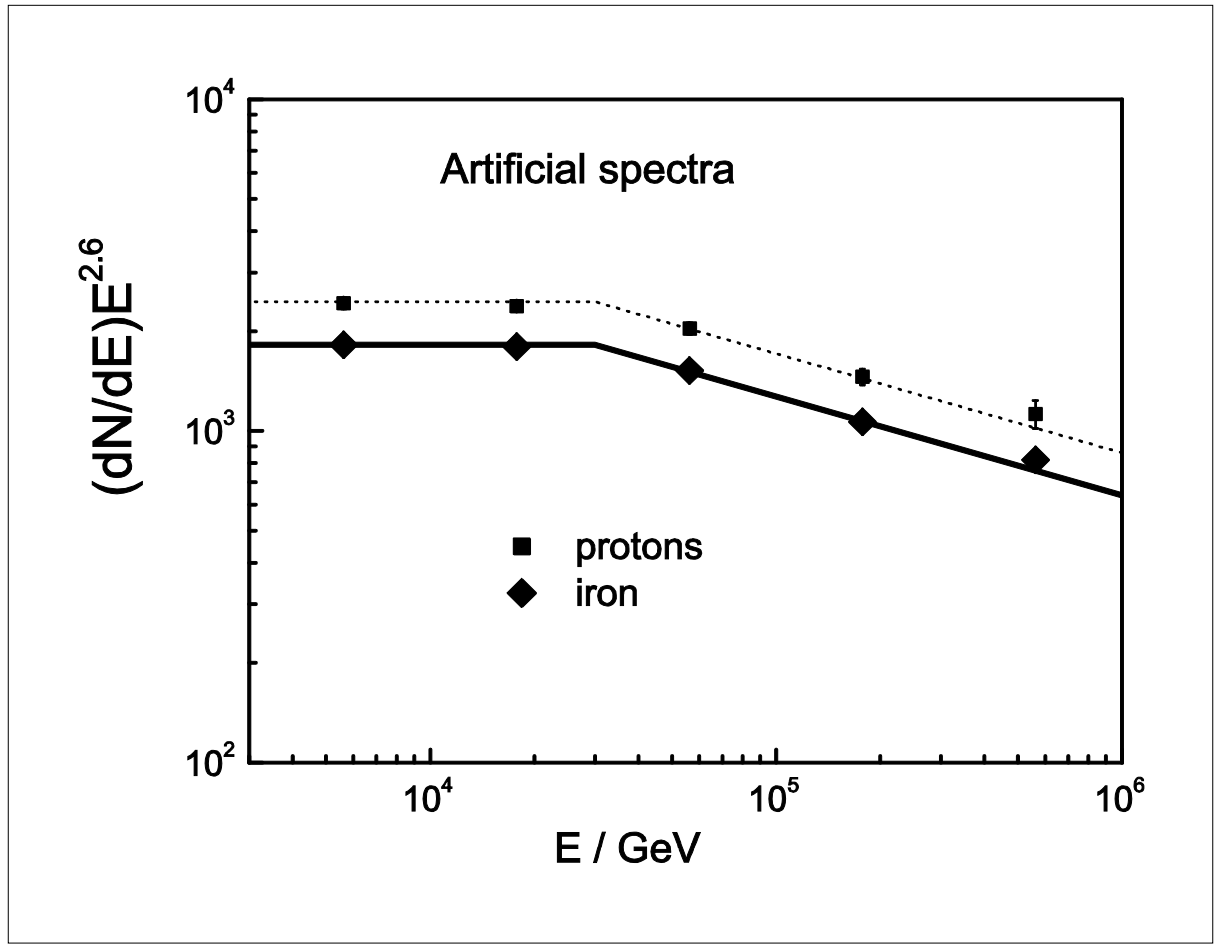

Figure 9. The simulated (lines) and reconstructed (points) energy spectra

Possible systematic uncertainties can be caused by different mechanisms. Electronic noise in silicon detectors can cause systematic uncertainty in the energy resolution. Low signals in detectors with magnitude less than 0.5 mip were neglected, in order to decrease this effect. The beam tests and simulation results were processed. The simulated and experimental reconstructed energy distributions are very close. The difference of mean reconstucted energies is near 4.6% [15]. It is significantly less than physical fluctuations.

Any uncertainty in energy measurements results in a systematic shift in the reconstructed flux intensity. This shift depends on the energy resolution too.

The registration efficiency was calculated as a generalized parameter and different effects were taken into account, including the resolution energy dependence. Thus the correction by the calculated registration efficiency allows restoration of energy spectra.

The systematic uncertainty of the flux depends on determinaton of the effective acceptance and the calibration of S. We evaluated these effects by variation of parameters. The systematic uncertainty of the effective acceptance is evaluated as 0.03 for heavy nuclei and 0.06 for protons due to acceptance dependence on the spectral exponent. The calibration of S is determined by periodical measurements of amplification factors for every channels. The systematic uncertainty of the calibration of S is equal to 0.0005.

The additional test of the KLEM technique is comparison of reconstructed spectra with spectra obtained by traditional ionisation calorimeter [27-30].

The energy spectra of different components of cosmic rays were reconstructed. The all particles' energy spectrum [28,29] was also reconstructed. The spectra of abundant components (protons, helium, carbon, oxygen, neon, magnesium, silicon, iron nuclei) are presented in figs. 10-17. The spectra for the KLEM technique and the ionization calorimeter (IC) [27] are shown. The KLEM and calorimeter data are almost independent because energy measurements are based on different detectors and different methods. Moreover the geometric factor for the calorimeter is significantly less than for the KLEM detector. Only about a quarter of events registered by the KLEM detector is registered also by IC. Therefore errors for the KLEM technique are reduced in comparison to the IC. The NUCLEON spectra are compared with results of different experiments (ATIC [1,2], CREAM [3,4], TRACER [5], AMS02 [6,7], SOKOL [8]).

The all particles spectra are presented in fig.18 in comparison with space and EAS experiments.

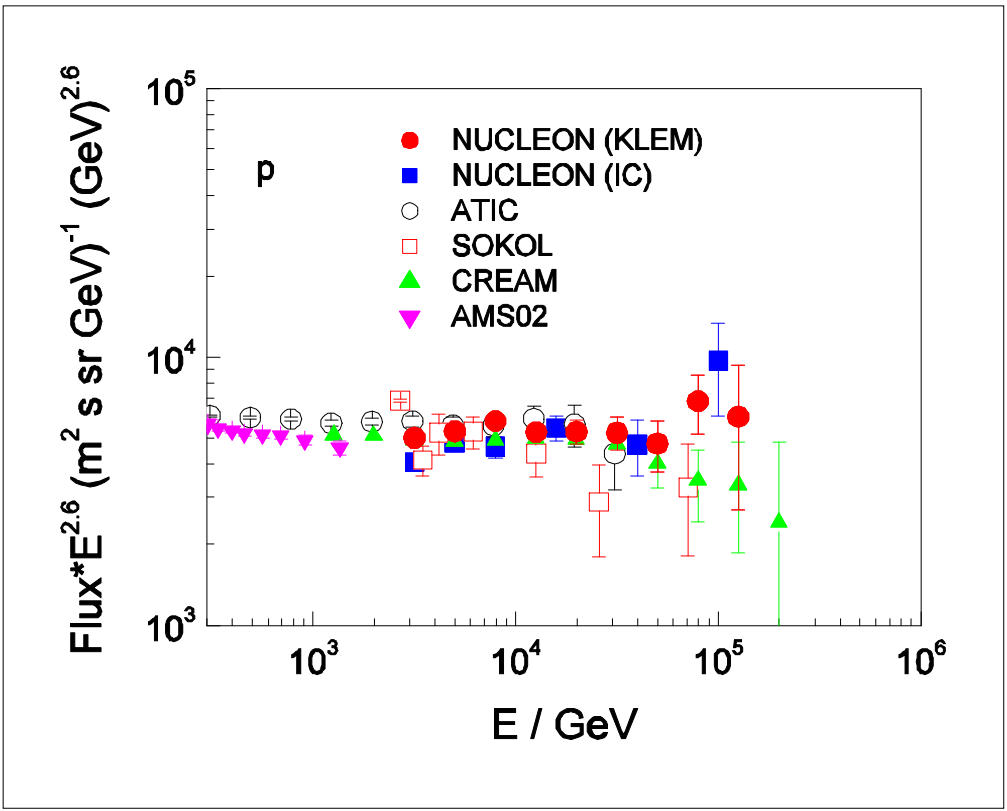

Figure 10. Proton spectrum

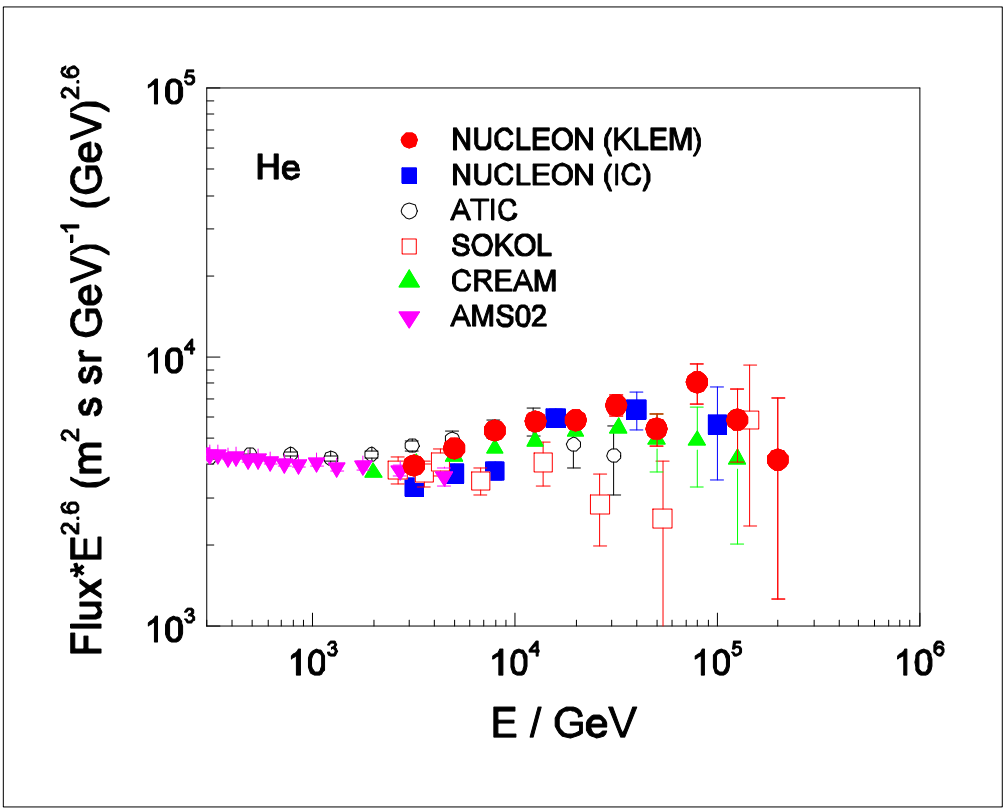

Figure 11. Helium spectrum

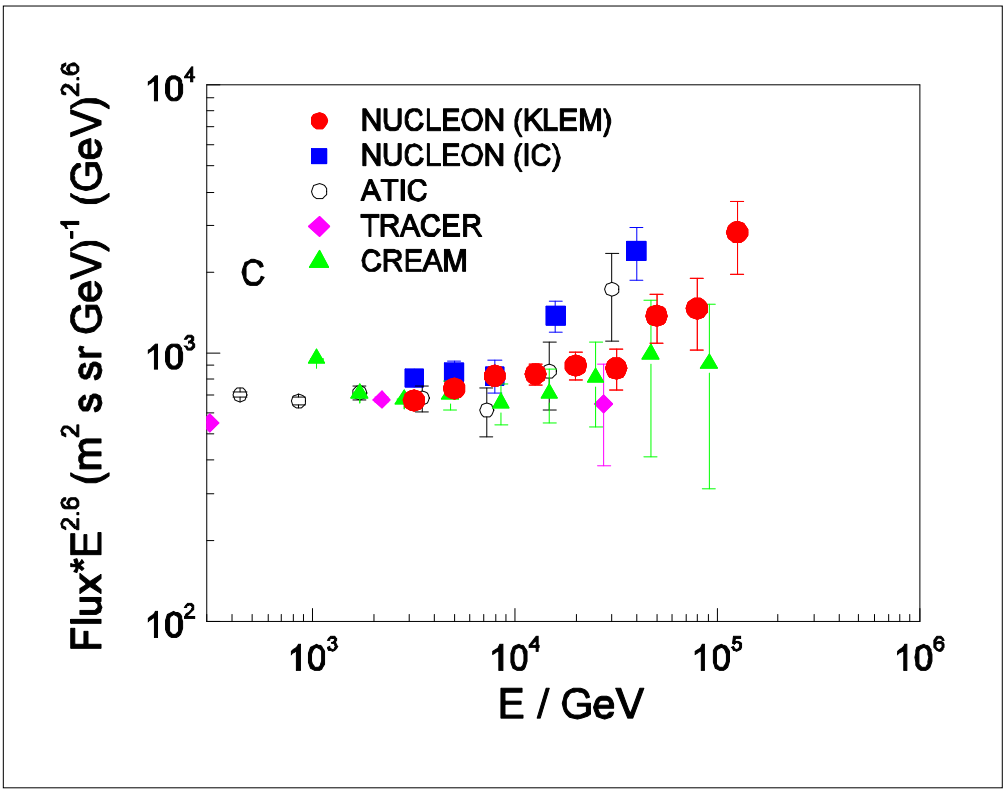

Figure 12. Carbon spectrum

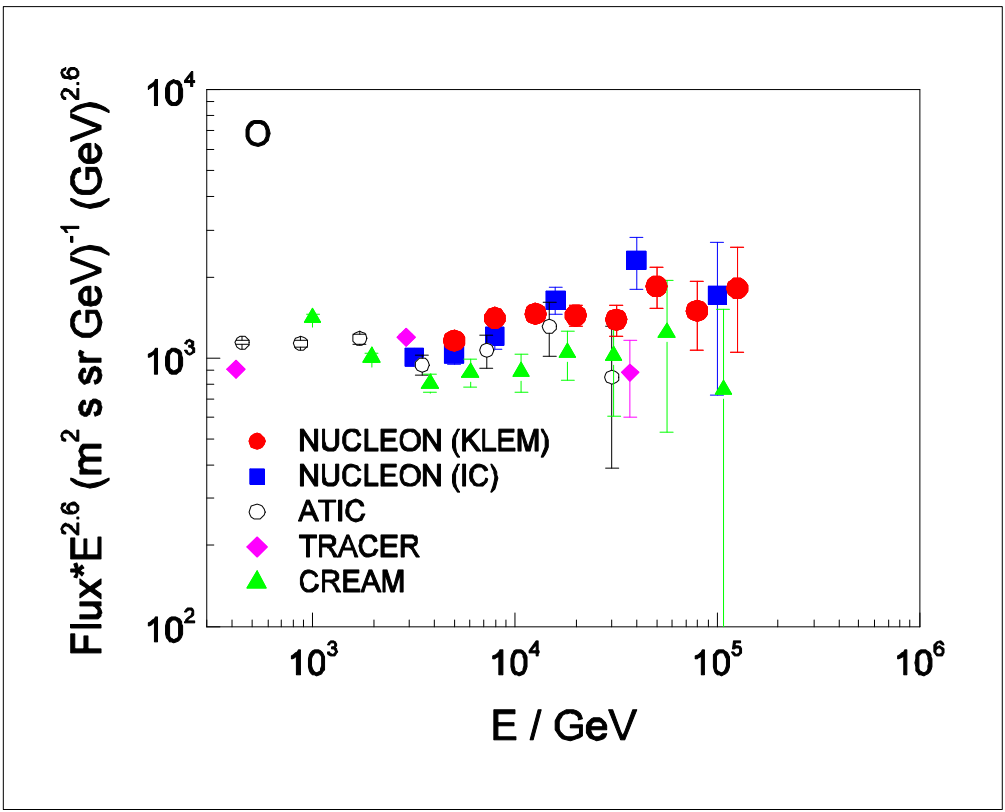

Figure 13. Oxygen spectrum

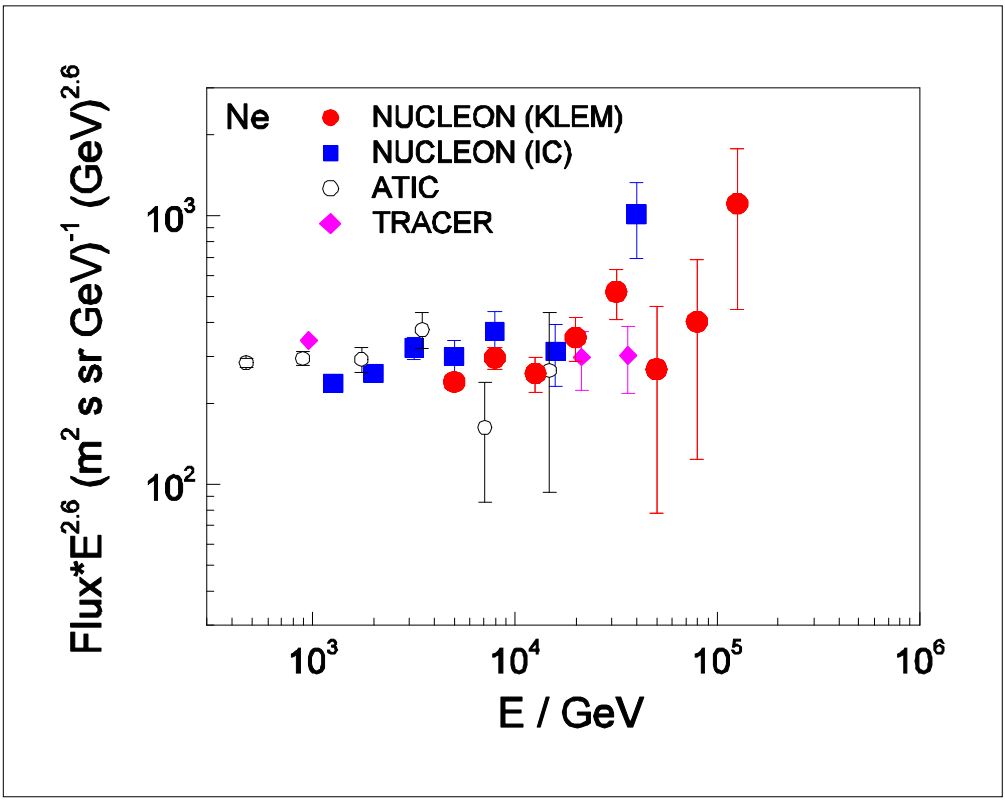

Figure 14. Neon spectrum

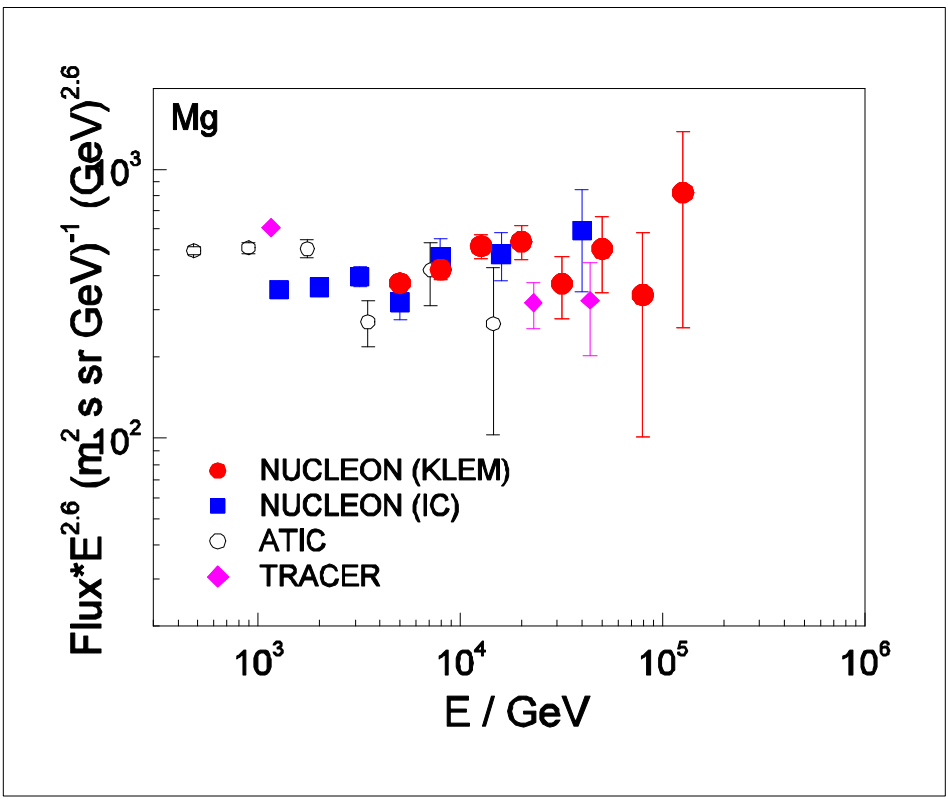

Figure 15. Magnesium spectrum

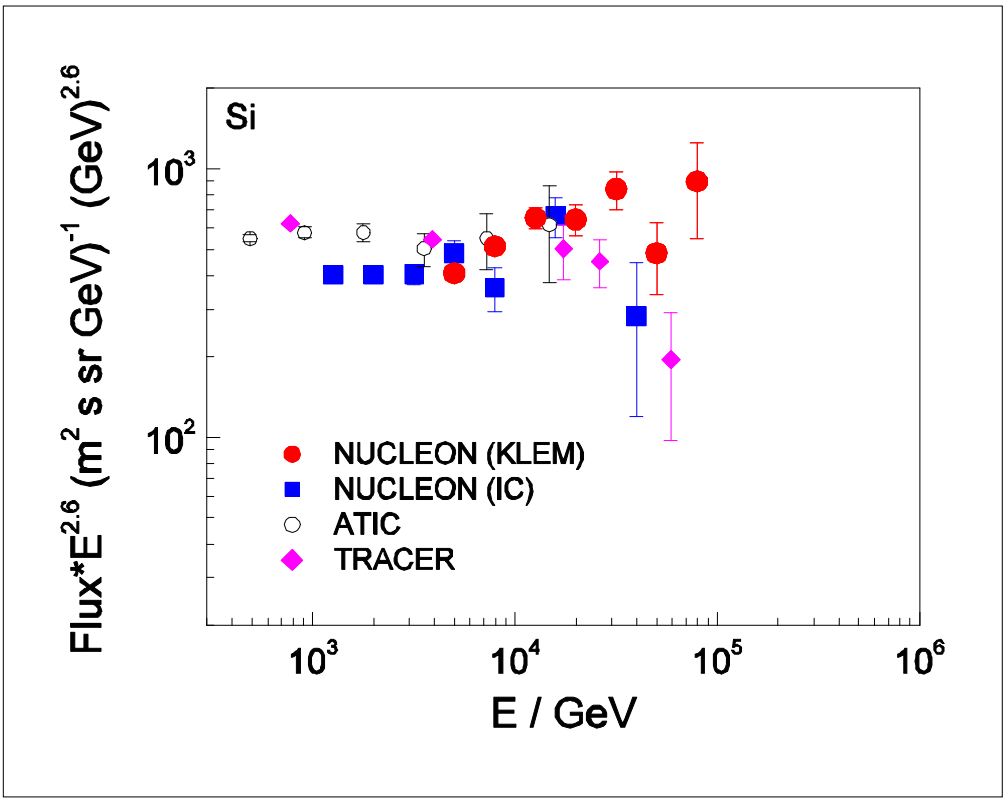

Figure 16. Silicon spectrum

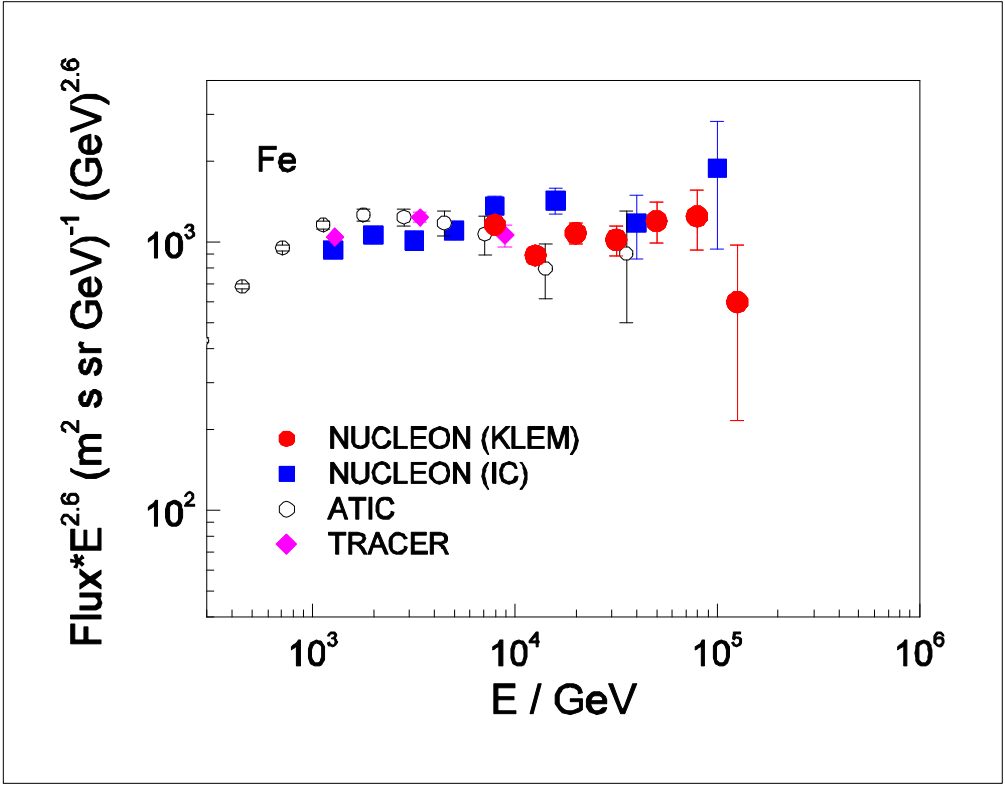

Figure 17. Iron spectrum

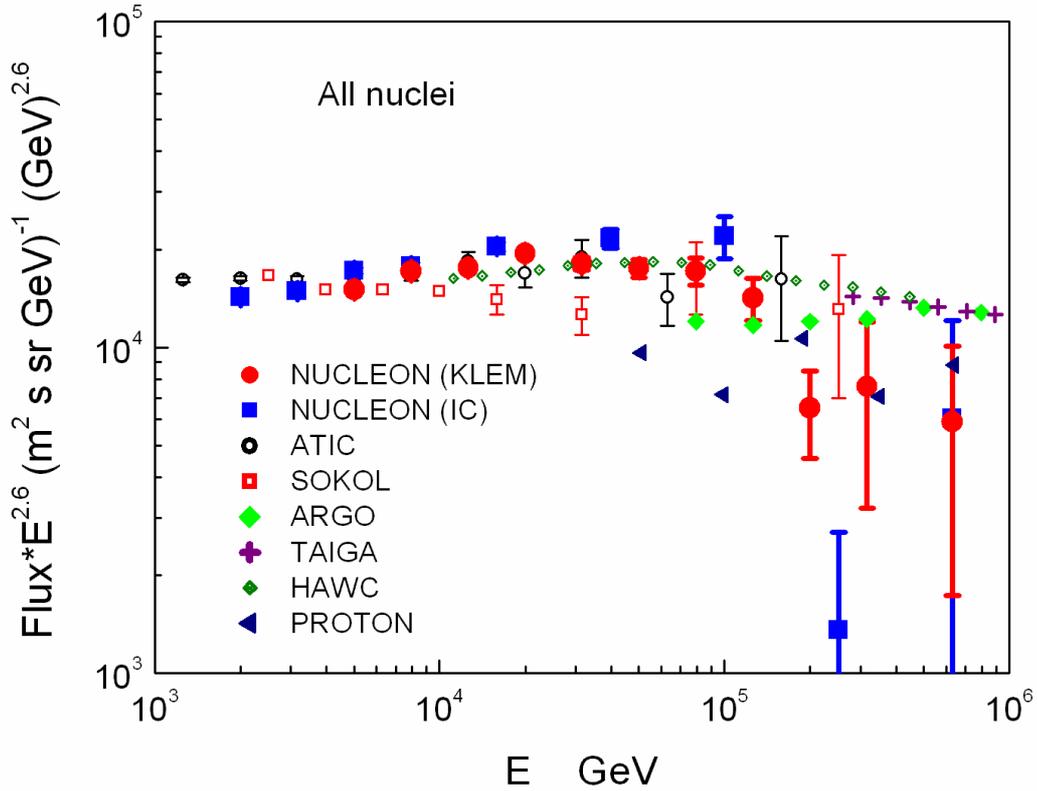

Figure 18. All particles spectrum

## 7. Discussion and conclusions

We can see some spectral peculiarities energies ~10 TeV. It can be interpreted by general dependence on magnetic rigidity [31].

The obtained energy spectra show good consensus on two different techniques of energy measurements. Thus, operability of a new KLEM technique in the wide energy range is confirmed.

Comparison of the energy spectra obtained by the NUCLEON experiment, and the results of other experiments, show good similarity to the energy range studied previously. At the same time, the NUCLEON data stretch to the area of energies higher than 100 TeV/particle for abundant nuclei, where there are no other experiments or their statistical material is too small.

A remarkable hardening of spectra is observed; this can possibly be explained by the presence of a few local sources of cosmic rays [32].

**Acknowledgments**
We acknowledge support from the Russian Space Agency (RosCosmos), Russian Academy of Sciences (RAS), JSC SRC Progress. The reported study was supported by the Supercomputing Center of Lomonosov Moscow State University [33].